\documentclass[twoside,twocolumn,9pt]{article}
\usepackage{extsizes}
\usepackage[super,sort&compress,comma]{natbib} 
\usepackage[version=3]{mhchem}
\usepackage[left=1.5cm, right=1.5cm, top=1.785cm, bottom=2.0cm]{geometry}
\usepackage{balance}
\usepackage{mathptmx}
\usepackage{sectsty}
\usepackage{graphicx} 
\usepackage{lastpage}
\usepackage{amssymb}
\usepackage[format=plain,justification=justified,singlelinecheck=false,font={stretch=1.125,small,sf},labelfont=bf,labelsep=space]{caption}
\usepackage{float}
\usepackage{fancyhdr}
\usepackage{fnpos}
\usepackage[english]{babel}
\addto{\captionsenglish}{%
  
}
\usepackage{array}
\usepackage{droidsans}
\usepackage{charter}
\usepackage[T1]{fontenc}
\usepackage[usenames,dvipsnames]{xcolor}
\usepackage{setspace}
\usepackage[compact]{titlesec}
\usepackage{hyperref}
\usepackage{placeins}

\usepackage{epstopdf}

\definecolor{cream}{RGB}{222,217,201}

\begin{document}

\pagestyle{fancy}
\thispagestyle{plain}
\fancypagestyle{plain}{
\renewcommand{\headrulewidth}{0pt}
}

\makeFNbottom
\makeatletter
\renewcommand\LARGE{\@setfontsize\LARGE{15pt}{17}}
\renewcommand\Large{\@setfontsize\Large{12pt}{14}}
\renewcommand\large{\@setfontsize\large{10pt}{12}}
\renewcommand\footnotesize{\@setfontsize\footnotesize{7pt}{10}}
\makeatother

\renewcommand{\thefootnote}{\fnsymbol{footnote}}
\renewcommand\footnoterule{\vspace*{1pt}%
\color{cream}\hrule width 3.5in height 0.4pt \color{black}\vspace*{5pt}} 
\setcounter{secnumdepth}{5}

\makeatletter 
\renewcommand\@biblabel[1]{#1}            
\renewcommand\@makefntext[1]%
{\noindent\makebox[0pt][r]{\@thefnmark\,}#1}
\makeatother 
\renewcommand{\figurename}{\small{Fig.}~}
\sectionfont{\sffamily\Large}
\subsectionfont{\normalsize}
\subsubsectionfont{\bf}
\setstretch{1.125} 
\setlength{\skip\footins}{0.8cm}
\setlength{\footnotesep}{0.25cm}
\setlength{\jot}{10pt}
\titlespacing*{\section}{0pt}{4pt}{4pt}
\titlespacing*{\subsection}{0pt}{15pt}{1pt}

\fancyfoot{}
\fancyfoot[LO,RE]{\vspace{-7.1pt}\includegraphics[height=9pt]{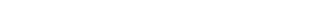}}
\fancyfoot[CO]{\vspace{-7.1pt}\hspace{13.2cm}\includegraphics{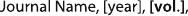}}
\fancyfoot[CE]{\vspace{-7.2pt}\hspace{-14.2cm}\includegraphics{head_foot/RF}}
\fancyfoot[RO]{\footnotesize{\sffamily{1--\pageref{LastPage} ~\textbar  \hspace{2pt}\thepage}}}
\fancyfoot[LE]{\footnotesize{\sffamily{\thepage~\textbar\hspace{3.45cm} 1--\pageref{LastPage}}}}
\fancyhead{}
\renewcommand{\headrulewidth}{0pt} 
\renewcommand{\footrulewidth}{0pt}
\setlength{\arrayrulewidth}{1pt}
\setlength{\columnsep}{6.5mm}
\setlength\bibsep{1pt}

\makeatletter 
\newlength{\figrulesep} 
\setlength{\figrulesep}{0.5\textfloatsep} 

\newcommand{\topfigrule}{\vspace*{-1pt}%
\noindent{\color{cream}\rule[-\figrulesep]{\columnwidth}{1.5pt}} }

\newcommand{\botfigrule}{\vspace*{-2pt}%
\noindent{\color{cream}\rule[\figrulesep]{\columnwidth}{1.5pt}} }

\newcommand{\dblfigrule}{\vspace*{-1pt}%
\noindent{\color{cream}\rule[-\figrulesep]{\textwidth}{1.5pt}} }

\makeatother

\twocolumn[
  \begin{@twocolumnfalse}
{\includegraphics[height=30pt]{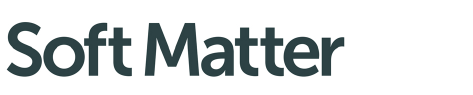}\hfill\raisebox{0pt}[0pt][0pt]{\includegraphics[height=55pt]{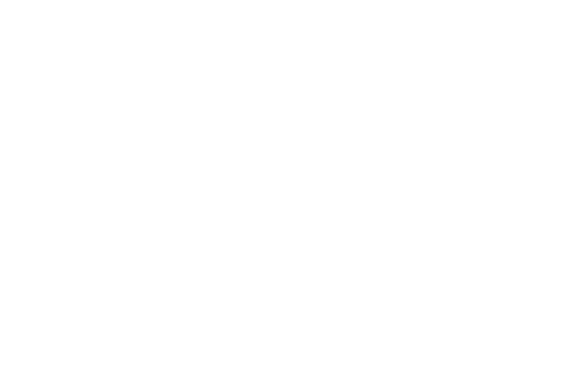}}\\[1ex]
\includegraphics[width=18.5cm]{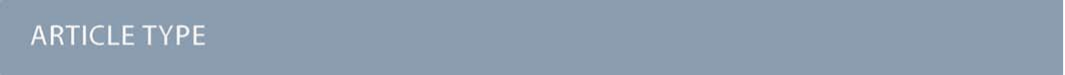}}\par
\vspace{1em}
\sffamily
\begin{tabular}{m{4.5cm} p{13.5cm} }

\includegraphics{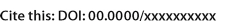} & \noindent\LARGE{\textbf{Rifts in Rafts$^\dag$}} \\
\vspace{0.3cm} & \vspace{0.3cm} \\

 & \noindent\large{Kh\'a-\^I T\^o$^{\ast}$\textit{$^{a}$}, and Sidney R. Nagel\textit{$^{a}$}} \\

\includegraphics{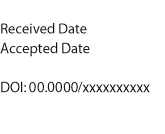} & \noindent\normalsize{
A particle raft floating on an expanding liquid substrate provides a macroscopic analog for studying material failure.  The time scales in this system allow both particle-relaxation dynamics and rift formation to be resolved.  In our experiments, a raft, an aggregate of particles, is stretched uniaxially by the expansion of the air-liquid interface on which it floats. Its failure morphology changes continuously with pulling velocity. This can be understood as a competition between two velocity scales: the speed of re-aggregation, in which particles relax towards a low-energy configuration determined by viscous and capillary forces, and the difference of velocity between neighboring particles caused by the expanding fluid. This competition selects the cluster length, i.e., the distance between adjacent rifts.  A model based on this competition is consistent with the experimental failure patterns.
} \\

\end{tabular}

 \end{@twocolumnfalse} \vspace{0.6cm}

  ]

\renewcommand*\rmdefault{bch}\normalfont\upshape
\rmfamily
\section*{}
\vspace{-1cm}


\footnotetext{\textit{$^{a}$~The Department of Physics and the James Franck and Enrico Fermi Institute\, The University of Chicago\, IL 60637.  E-mail: cytao@uchicago.edu}}

\footnotetext{\dag~Electronic Supplementary Information (ESI) available: [details of any supplementary information available should be included here]. See DOI: 10.1039/cXsm00000x/}

\section{Introduction}
The failure of a sheet of material pulled from its two opposing edges has often been characterized as being either brittle, where a thin crack propagates rapidly across the material breaking it into two, or ductile, where plastic deformation causes the material to deform, neck and eventually break. 
A number of reviews have focused on fracture in these regimes separately.  For ideally brittle solids, the emphasis has been on how stresses are concentrated to a ``process zone'', in which damage occurs ahead of the propagating crack while the rest of the solid remains in the elastic regime~\cite{FINEBERG19991, doi:10.1146/annurev-conmatphys-070909-104019}. The reviews of ductile flow, particularly in soft amorphous materials, have emphasized the role of extensive and collective plastic events in producing global deformation~\cite{doi:10.1146/annurev-conmatphys-062910-140452, RevModPhys.90.045006}.
While these reviews approach the topic of material failure from these opposite perspectives, they acknowledge that the brittle/ductile dichotomy is oversimplified due to the highly complex nature of the phenomena. In some systems, the spatial distribution of plastic events can be tuned continuously from the atomic scale (as in an ideally brittle crack) to the system size without significant changes in material shape. Disorder is an important factor in determining the nature of this failure zone~\cite{Roux1988, PhysRevB.76.144201, PhysRevLett.100.055502, PhysRevLett.110.185505}.

Material rigidity is also found to be a control parameter for the creation of wide failure zones~\cite{Driscoll10813}; as the rigidity is reduced, the width of the failure zone diverges 
and the crack-propagation speed decreases by several orders of magnitude.
In those experiments and simulations, no rearrangement or bond creation was allowed once a bond was broken. Thus, because there was no rearrangement or change in material shape, these systems cannot be described as ductile. 

In this paper, we investigate the failure during expansion of a particle raft composed of sub-millimeter particles floating at an air-liquid interface. 
This is a particularly interesting system because (i) the particles are macroscopic, so that their individual motions can be resolved to allow direct observation of the relaxation, (ii) the pulling speed can be varied over several orders of magnitude, (iii) the particles are coupled to a linearly deforming liquid substrate which does not store stress, (iv) as in the previous example,
the failure can be distributed throughout the material, 
and most importantly (v) the particles can rearrange and find new neighbors as the rifts evolve. These systems are therefore distinct from those mentioned above: like them the failure occurs throughout the material but instead of simply breaking bonds, 
plastic deformation and rearrangement occurs throughout the raft.

Our experiment reveals that 
the expansion speed controls the structural morphology of the failure. 
As the speed increases, new rifts form as the raft breaks up into ever-finer structures until, at the highest strain rates, the structures reach the individual particle level. We determine that this behavior is a competition between the pulling speed and the microscopic dynamics of particle relaxation and have developed a one-dimensional linear (in)stability analysis which provides a good description of this behavior.

Previous studies of granular rafts have focused on characterizing their elasticity and buckling behavior under compression~\cite{PhysRevE.86.031402, PhysRevMaterials.1.042601, PhysRevLett.102.138302, Vella_2004, doi:10.1021/acs.langmuir.5b01652}. 
A few papers have examined rafts under tensile stress and have observed localized fracture events by introducing a radial gradient in the flow field~\cite{Bandi_2011, peco2017influence, KIM201954}. In the quasi-static regime, the ductile behavior of particle rafts shows a dependence on particle size at small strains~\cite{D0SM00839G}. However, the dependence of the failure morphology on the pulling speed was not examined. 

In the case of bubble rafts, a failure pattern as a function of strain rate and system size was investigated but the distributed-failure regime was not accessed~\cite{ARCINIAGA201136}. In that case, the observed change in failure morphology was simply ascribed to a transition between brittle and ductile behavior.  By exploring wider dynamical ranges in both strain and strain rate, we indicate that this is, perhaps, an oversimplified interpretation. 
We observe a change in failure morphology as rifts, distributed homogeneously, form throughout the raft; the distance between the rifts changes continuously with pulling speed.  We interpret this in terms of a single form of failure that is governed by expansion rate. 

Another related system is crack formation in a thin material sheet on a \textit{solid} substrate. These include nanoparticle films on expanding polymer membranes~\cite{C4FD00243A, LU20101679} and drying colloidal monolayers on glass~\cite{skjeltorp1988fracture, Routh_2013}. In both cases, the systems exhibit cracking patterns throughout the material but with no obvious dependence on the pulling or drying speed. These patterns are governed by the interactions with the underlying solid substrate; the relaxing elastic stress in the thin sheet competes with the interaction holding the sheet to the substrate. In contrast, the system we study is a particle raft on a \textit{liquid} substrate which does not store any shear stress. The mechanism determining the pattern of crack formation in these rafts is therefore distinct from what occurs on solid substrates. 

In our experiment, the affine expansion of the liquid surface on which the particles float can be thought of as an expanding metric for the particle positions. In contrast to the short-range repulsion between granular matter, particles in floating rafts also have a longer-range attraction due to lateral capillary forces. These vary inversely with the particle separation asymptotically~\cite{paunov1993lateral, vella2005cheerios, PhysRevE.83.051403}, which has the same form as two-dimensional gravitational attraction~\cite{18717320050101}. We therefore note that this situation bears a resemblance to structure formation during cosmological expansion of the universe~\cite{doi:10.1146/annurev-astro-081811-125502}. Our experiment, which measures the cluster formation in the two-dimensional raft as a function of uniaxial pulling speed, is similar to structure formation in three dimensions but with overdamped rather than underdamped dynamics and on obviously much smaller scales.

\section{Experimental apparatus and methods}\label{sec2}
\begin{figure}[!htbp]
\centering
\includegraphics[width=.75\linewidth]{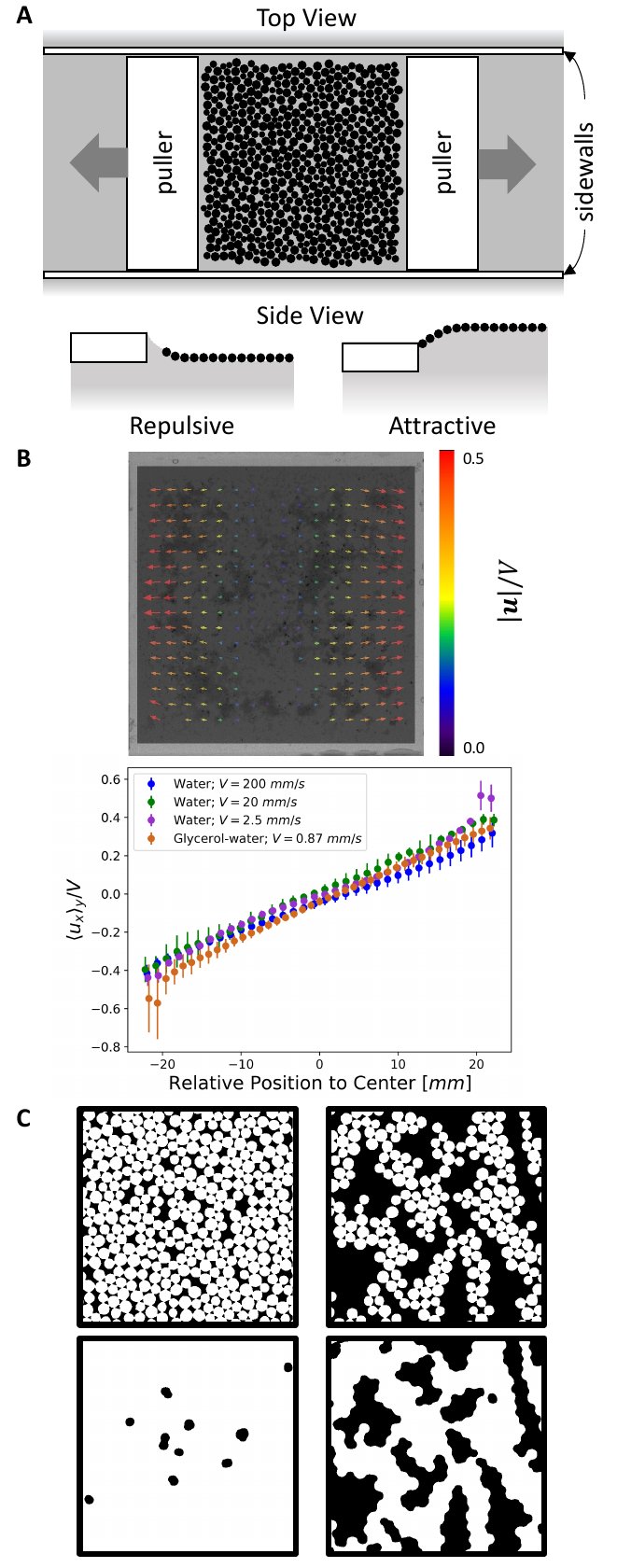}
\caption{(A) Top: A schematic of the experimental apparatus shows the raft in the center with two pullers on each side that extend the raft along one axis at a fixed pulling velocity, $V$. Bottom: A side view with either repulsive or attractive boundary conditions. 
(B) The Particle Image Velocimetry (PIV) result of the two pullers moving in the opposite directions on air-liquid interfaces with different $Re$.Top: The PIV map for an air-water interface with $V=20$ $mm/s$ at the onset of pulling. The color bar shows the speed $|\mathbf{u}|$ normalized by $V$. Bottom: For all $Re$, the average velocity parallel to the pulling direction increases linearly with respect to the horizontal position. (C) Example of procedure to remove the spacings between particles that touch each other even in a compact raft. Left and right rows show a packing before and after expansion. In each case the image on the top is before image processing and on the bottom is after image processing.}
\label{fig1new}
\end{figure}

A schematic of the experimental apparatus is shown in Fig.~\ref{fig1new}A. We use rafts comprised of spherical polyethylene particles floating at an air-liquid interface. We use both deionized water and a $60$ $w/w \%$ glycerol-water mixture as underlying fluids with different viscosities.  Polydisperse packings are made by mixing approximately equal volume of particles chosen with two diameter ranges: $d=550 \pm 50$ $\mu m$ (small) and $d=655\pm 55$ $\mu m$ (large). These submillimeter particles coalesce into a floating raft due to the lateral capillary attraction (known as the “Cheerios” effect~\cite{paunov1993lateral,vella2005cheerios,PhysRevE.83.051403}) between the particles.  The rafts have initial packing fraction of $73\pm 1\%$.

The raft is enclosed by four boundaries. The initial lengths parallel, $L_{x0}$, and perpendicular, $L_{y0}$, to the direction of pulling are $L_{x0} \approx L_{y0} \approx 53$ $mm$.
Two polypropylene plates with sharp edges are used to pull the raft apart. They are held with their bottom surfaces just below the air-liquid interface. As the menisci are pinned at the pullers’ edges, the boundaries can be made either hydrophilic or hydrophobic by adjusting the water level slightly. As shown in the bottom of Fig.~\ref{fig1new}A, for a hydrophilic (hydrophobic) surface, the meniscus points upward (downward) and creates a repulsive (attractive) interaction between the particles and the pulling boundary. When the boundary is repulsive (hydrophilic), it does not contact the raft directly; the expansion of the liquid surface, on which the particles float, creates an affine expansion of the underlying metric along the direction of the pulling velocity. This causes the raft to expand in that direction. When the pullers are attractive, in addition to the expansion of the liquid surface there is also the pulling of the raft by the moving walls to which the raft is connected. Two hydrophilic acrylic sidewalls are placed along the direction of pulling to keep the particle rafts from touching those side walls as the raft is pulled. This reduces the friction while keeping the raft confined.

We measure the velocity field of the expanding liquid surface and find that the surface on which the particles float expands uniformly over time in the pulling direction. We have studied both repulsive and attractive boundary conditions for the pullers. The data presented here will be for repulsive conditions. In the Supplementary Information, we show movies using attractive boundary conditions that show similar behavior as that presented in the main text.

The two pullers are moved in opposite directions as shown in Fig.~\ref{fig1new}A. Each one is moved at a constant speed $V/2$. This allows the center of the raft to remain fixed in the laboratory frame of reference. 
We vary $V$ in different experiments over several decades: from $2.5$ $mm/s$ to $200$ $mm/s$ for deionized water and from $0.87$ $mm/s$ to $42$ $mm/s$ for the glycerol-water mixture.
The characteristic velocity difference between two neighboring particles is $V/(N_x-1) \approx V/N_x$, where $N_x=L_{x0}/\langle d \rangle$ is the number of particles in the pulling direction. The Reynolds number, $Re= \rho d(V/N_x)/\eta$, ranges from $7.2 \times 10^{-3}$ to $8.1 \times 10^{-1}$ for water and $2.8\times10^{-4}$ to $1.9\times10^{-2}$ for glycerol-water mixture. We do not go faster than $200$ $mm/s$ with water because when pulling at high velocity, we observe significant surface waves that affect our measurements. For the glycerol-water experiments, we go to the slowest possible speed that our motor can reliably control. Since $L_x=L_{x0}+Vt$, the liquid strain rate across the system
\begin{equation}\label{eq1}
    \dot{\varepsilon} = \frac{V}{L_x} = \frac{V}{L_{x0}+Vt}
\end{equation}
decreases monotonically with time. The initial strain rate,$\dot{\varepsilon}_0 = V/L_{x0}$, was varied from $1.6 \times 10^{-4}$ $s^{-1}$ to $3.7$ $s^{-1}$. The experiments stop at a maximum liquid strain value $\varepsilon_{max} \approx 1.8$.

\textit{\textbf{Fluid and particle properties:}}  For our low viscosity underlying fluid, we use deionized water with density $\rho_w=998$ $kgm^{-3}$, dynamic viscosity $\eta_w= 0.95$ $mPas$ and surface tension $\sigma_w= 0.073$ $Nm^{-1}$ at $22^{\circ}C$.  For the higher viscosity fluid, we use a $60$ $w/w\%$ glycerol-water mixture with density $\rho_{gw}=1150$ $kgm^{-3}$, dynamic viscosity $\eta_{gw}=10.1$ $mPas$, and surface tension $\sigma_{gw}=0.064$ $Nm^{-1}$ at $22^{\circ}C$.  For the deionized water experiments, the small and large particles have densities $1025$ and $1080$ $kgm^{-3}$ respectively. For the experiments on the glycerol-water fluid, both small and large particles have density $1130$ $kgm^{-3}$.  


\textit{\textbf{Particle Image Velocimetry:}}  To measure the expansion of the liquid surface due to the extension of the pullers, we perform Particle Image Velocimetry (PIV) measurements over our entire experimental range of Re. The results are shown in Fig.~\ref{fig1new}B. We spread a dilute layer of very light, non-interacting floating particles on the liquid surface prior to pulling. By tracking the motion of these particles and computing the correlation between adjacent frames, we determine that the underlying fluid flows lead to an affine expansion of the surface; the spacing of the particles in the direction of the pulling increases linearly in time. The PIV map of nearly the entire surface for air-water interface with $V=20$ $mm/s$ is shown in the top of Fig.~\ref{fig1new}B. The particle velocities are increase uniformly in the direction of pulling. The bottom of Fig.~\ref{fig1new}B shows the gradient of the velocity is constant in the direction of pulling, which is valid across the entire range of $Re$ in our experiments. We do observe larger deviations from the affine expansion when the particles are closer to the pullers, especially at a later time. At low $Re$, the non-affine flow is dominated by the secondary currents flowing in from the slits between the pullers and the sidewalls. At high $Re$, the surface wave in the third dimension is the main contributor to the non-affine expansion.


\textit{\textbf{Image Recording:}}  The motions of the particles are recorded using Phantom VEO 640S and Proscilica GX3300 cameras. The frame rates are adjusted so that the displacement of a particle between frames is no larger than $d_{max}$.  In our experiments, we first tried measuring the pair-correlation function, $g(\mathbf{r})$, and the structure factor, $S(\mathbf{k})$, using images captured by a high-resolution camera. However, the results did not show clear signatures of the rift formation and cluster sizes. (See Supplementary Information for the results.)  We therefore measured directly the size of the clusters between adjacent rifts in a raft. 

\textit{\textbf{Image Processing:}}  To obtain this information, we do not want to include the microscopic holes between the spherical particles that would be there even when the raft is closely packed. 
The original image of the raft is thresholded and binarized so that the particles are white and the background is black.
The radius of each particle is increased so that the small gaps between particles, due to the fact that spheres cannot tile space fully, are no longer present. The perimeters of the new clusters are then decreased by the same amount as they were originally expanded. This closes the gaps internal to the raft but leaves only the gaps or rifts that are formed due to raft expansion. 
This morphological operation to the binarized images results in an increase in the area of the particles. The size of the dilation kernel is determined by the square root of the average hole area in the initial packing. A reverse morphological operation (erosion) is then performed to shrink the clusters around their perimeters while leaving the particles swelled on the interior. This procedure removes the holes between particles that are comparable to the spacing in the original packings so that only the small newly-created cracks, the rifts due to the raft expansion, remain. This is shown in Fig.~\ref{fig1new}C. To measure the cluster lengths in the horizontal direction, we take one-pixel height horizontal slices of the processed images and discard short slices at the left and right edges of the rafts to reduce noise in our measurement.

\begin{figure*}[!htbp]
 \centering
 \includegraphics[width=.8\linewidth]{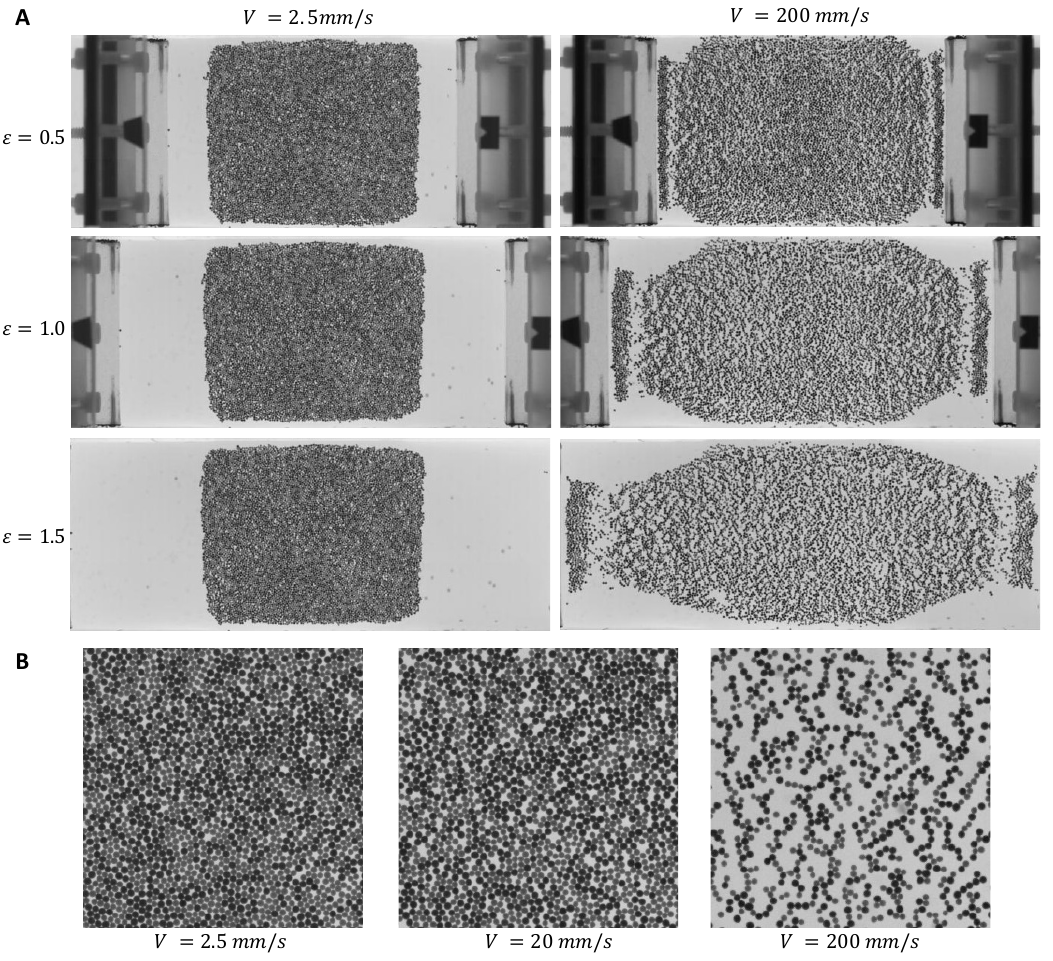}
 \caption{Failure morphology at different pulling speeds.  
 (A) Snapshots of air-water experiments at different velocity, $V$, using repulsive boundary conditions.  The series of images show the expansion of the rafts at low ($V=2.5$ $mm/s$) and high ($V= 200$ $mm/s$) velocities. (B) Zoomed-in images show structure in the bulk of the rafts for a strain of $\varepsilon =1.5$ at  $V=2.5, 20, 200$ $mm/s$ respectively. }
 \label{fig2new}
\end{figure*}
\section{Results}
\subsection{Experiment}
As the pulling boundaries are separated, there is a change in the morphology of the raft as micro-cracks, or rifts, begin to form as shown in Fig.~\ref{fig2new}A.  The two columns correspond to different pulling speeds, $V$.  The images in each column are taken from a single movie at equally spaced values of the liquid strain, $\varepsilon$.

As the raft expands in the horizontal direction, the rifts become larger, as can be seen in subsequent images in each column. These small cracks, or rifts, are distributed diffusely throughout the entire system.
The horizontal distance between adjacent rifts determines the cluster length, $\ell$.

The morphology depends strongly on the pulling speed, $V$, as seen in the difference between the left and right columns of Fig.~\ref{fig2new}A. The number of micro-cracks and $\ell$, both depend strongly on $V$. When $V$ is small enough, the raft is only slightly sheared although the underlying liquid has been stretched by a factor of $2.5$ (corresponding to $\varepsilon=1.5$). After the initial disturbance, the raft remains unchanged for the rest of the expansion; no significant micro-cracks can be observed and the cluster width $\ell$ remains close to the initial system size, $L_{x0}$.

With increasing $V$, the number of rifts increases while the cluster size, $\ell$, decreases. When $V$ is increased further, the raft stretches along the pulling direction but
remains essentially unchanged in the transverse direction except for the corners, which are disturbed by the secondary flow for the less-viscous fluid (deionized water). In this regime, it behaves like a sheet with zero Poisson’s ratio. As $V$ increases, $\ell$ decreases until it approaches the size of a single particle, as shown in the right column of Fig.~\ref{fig2new}A. Figure~\ref{fig2new}B shows how the internal features change with increasing $V$: a relatively close-packed structure gradually breaks up into clusters which are only a few, or sometimes only one, particles in width.

\begin{figure}[!htbp]
\centering
\includegraphics[width=.8\linewidth]{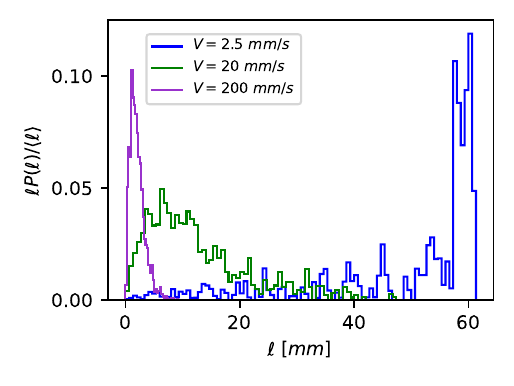}
\caption{Distribution of cluster lengths, $\ell$, for different velocity at fixed strain in air-water experiments. The most dominant length, that is the probability of cluster length $\ell$ multiplied by $\ell$ normalized by $\langle \ell \rangle = \Sigma\ell P(\ell)$, is plotted versus $\ell$ at a fixed strain $\varepsilon=1.0$  for three pulling speeds, $V$: $2.5$ $mm/s$ (blue), $20$ $mm/s$ (green) and $200$ $mm/s$ (purple).}
\label{fig2}
\end{figure}

To quantify this observation, we measure the cluster length, $\ell$, of each raft as the width of the cluster parallel to the direction of pulling. 
The image processing protocols are described in Sec. \ref{sec2}. 
Figure~\ref{fig2} shows the distribution of cluster sizes at a liquid strain $\varepsilon=1.0$ for different pulling velocities. The results are largely independent of $\varepsilon$ for $0.5 \lesssim\varepsilon\lesssim 1.5$.  Once formed, a rift does not immediately collapse.  (However, in the long-time regime, not probed in this experiment, the capillary forces will eventually dominate the interparticle interaction because, with a constant pulling speed, $\varepsilon$ decreases as $L_x$ increases as shown in Eq.~\ref{eq1}.)
The ordinate is $\ell P(\ell)/\langle \ell \rangle$, that is the most dominant cluster length for each velocity, where
$P(\ell)$ is the probability of finding a cluster of length $\ell$ and the average cluster length $\langle \ell \rangle = \Sigma \ell P(\ell)$. 
Since the statistics of small $\ell$ is always higher than that of large $\ell$, the most dominant length shows how much material has cluster length $\ell$ and helps better identify the cluster distribution at different $V$. 
At small $V$, a large portion of the distribution remains close to the initial cluster size, $L_{x0}$. (The clusters can be larger than the $L_{x0}$ because the raft can still be slightly stretched at small $V$.) The distribution shifts to smaller $\ell$ as $V$ increases. At our highest pulling speed, $V$, most clusters have lengths between $d$ and $2d$.

\begin{figure}[!htbp]
\centering
\includegraphics[width=.8\linewidth]{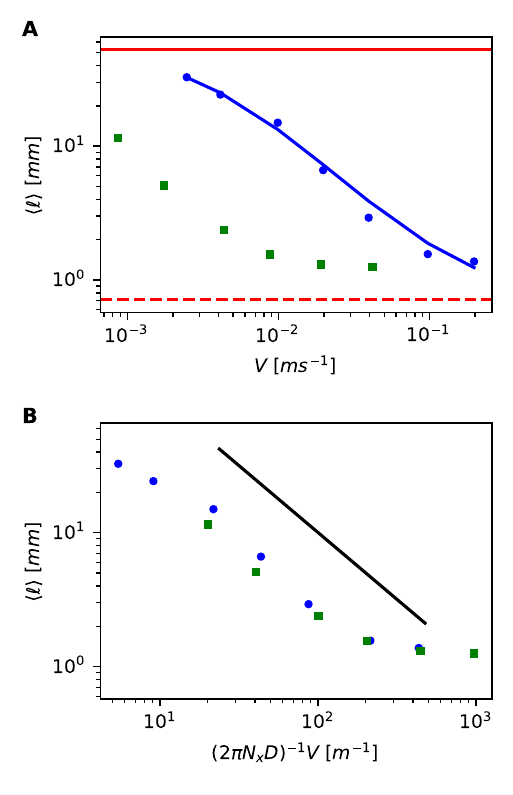}
\caption{Average cluster length, $\langle \ell \rangle = \Sigma\ell P(\ell)$, versus pulling speed, $V$ at fixed strain $\varepsilon=1.0$. (A) The blue and green points are measurements of $\langle \ell \rangle$ versus $V$ for deionized water and glycerol-water mixtures respectively. The blue curve is a fit of Eq.~\ref{eq2} to the water data set. The red solid and dashed lines indicate the initial system size $L_{x0}$ and largest particle diameter $d_{max}$ respectively. (B) The pulling speeds, $V$ , are scaled with $(2 \pi N_x D_w)^{-1}$ and $(2\pi N_x D_{gw})^{-1}$ derived from Eq.~\ref{eq6}. The experimentally measured values of $D_w$ and $D_{gw}$, the ratio between the lateral capillary force to the Stokes' drag, for the two fluids were used. The black line shows the theoretical prediction for water. }
\label{fig3}
\end{figure}

Figure~\ref{fig3} shows the average cluster length $\langle \ell \rangle$ versus pulling speed $V$ at a fixed value of the strain, $\varepsilon=1.0$ for both water and glycerol-water experiments. One can see that $\langle \ell \rangle$ decreases monotonically with increasing $V$ and saturates at high velocities. Because the length of a cluster cannot be significantly larger than the size of the system, $L_{x0}$, or smaller than a particle diameter, $d$, we interpolate between the two extremes in our air-water data using:

\begin{equation}\label{eq2}
\langle\ell\rangle = \frac{1}{aV^b+1/(L_{x0}-d_{max})}+d_{max}. 
\end {equation}

This fit is shown in Fig.~\ref{fig3}A.  Fitting to this form gives $b=1.1\pm0.2$ and $a = (1.2 \pm 0.4)\times 10^4$ in SI units.

\begin{figure}[!htbp]
\centering
\includegraphics[width=.8\linewidth]{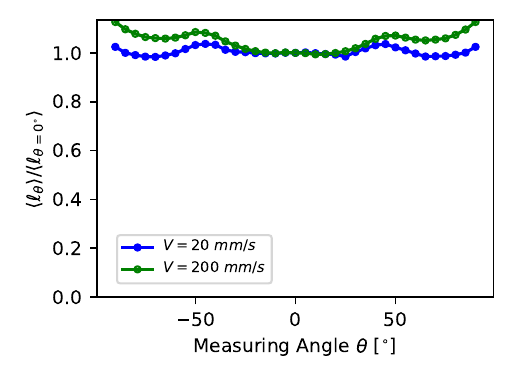}
\caption{Averaged cluster length measured at different measuring angles, $\langle\ell_{\theta}\rangle$, versus measuring angle, $\theta$, for air-water experiments at $V=20$ $mm/s$ and $200$ $mm/s$ at fixed strain $\varepsilon=1.0$. $\langle\ell_{\theta}\rangle$ is normalized by $\langle\ell_{\theta = 0^{\circ}}\rangle$, that is the averaged cluster length measured in the direction of pulling. We observe only small fluctuations of order 10\% in $\langle\ell_{\theta}\rangle/\langle\ell_{\theta = 0^{\circ}}\rangle$ with respect to $\theta$, indicating that the cluster orientation is fairly isotropic.}
\label{fig4}
\end{figure}

To characterize the structure and orientation of the clusters, we examine two air-water experiments with $20$ $mm/s$ and $200$ $mm/s$ at a fixed value of the strain, $\varepsilon = 1.0$. The average cluster lengths, $\langle \ell \rangle$, are $6.62$ $mm$ and $1.37$ $mm$, with percentage standard deviation of $97\%$ and $75\%$, respectively. The high variation in cluster length is mostly due to a wide distribution of the cluster orientations, as observed in the images Fig.~\ref{fig1new}B. To measure the orientation of the clusters, we define the measuring angle, $\theta$, which is the angle with respect to the pulling axis. We then compare the cluster length measured at different angles, $\ell_{\theta}$. Therefore, $\ell_{\theta = 0^{\circ}}$ is the cluster length measured horizontally. Figure~\ref{fig4} shows the average of $\ell_{\theta}$ normalized by $\langle\ell_{\theta = 0^{\circ}}\rangle$ versus measuring angle $\theta$ for these two experiments. We first observe that $\langle \ell_{\theta} \rangle$ at positive and negative angles are very symmetric, indicating that no significant asymmetrical forces were applied by the top or bottom boundaries. There are also some small features at $\theta = \pm 45^{\circ}$ and $90^{\circ}$, most apparent at $200$ $mm/s$.  However, this is an overall small effect at both speeds.
Therefore, while the clusters themselves are far from circular, their orientations are very isotropic.

\subsection{Analysis}
To understand the velocity-dependence of the failure morphology, we consider a one-dimensional chain of $N_x$ identical particles to model our rafts as they are pulled uniaxially in the $x$-direction. We assume that the dynamics are overdamped because $\rho d\dot{x}/\eta \ll 1$ (where $\dot{x}$ is the velocity of a particle, $\eta$ is the dynamic viscosity of the liquid bath, $d$ is the particle diameter and $\rho$ is the density of the liquid). 

At $t=0$, the particles are in contact with their neighbors with one end of the chain fixed and the other end pulled at a constant pulling speed $V$. The positions of the particles along the chain will remain evenly spaced and increase linearly in time due to the affine motion of the fluid. The coordinate of the $j^{th}$ particle is $x_j$ and the uniform distance between the centers of adjacent particles is $\Delta x$. The spacing between the particles increases as $\frac{d\Delta x}{dt}= V/(N_x-1) \approx V/N_x$.  We perturb the particles from their equilibrium positions by a small displacement, $u(x)$.

Each particle will feel the lateral capillary forces from its two neighbors which depends on the distance to those particles as calculated in Ref. \cite{vella2005cheerios}. For particle, $j$, the distance to its neighbor on the left is $\Delta x_{j,l}=\Delta x+u(x_j)-u(x_{j-1})$,  and to its neighbor on the right is $\Delta x_{j,r}= \Delta x+u(x_{j+1})-u(x_j)$. This analysis, although it appears to have the same simple form as a chain of balls connected by springs, is different in three crucial regards: (i) as we shall see, due to the nature of the interparticle potential, the chain is inherently unstable to any expansion because the effective spring constant decreases with interparticle distance; (ii) this leads to a relaxation velocity that depends on cluster size and which then competes with the expansion velocity and (iii) the particle motions are overdamped and their motion obeys Stokes drag. We obtain:
\begin{equation}\label{eq3}
   3\pi \eta d \alpha \frac{du(x_j)}{dt}  = \pi\sigma d B_o^{5/2}\Sigma^2\Big(-K_1(\frac{\Delta x_{j, l}}{L_c})+K_1(\frac{\Delta x_{j, r}}{L_c})\Big)
\end{equation}
where $\eta$ and $\sigma$ are the dynamic viscosity and surface tension of the liquid respectively. $\alpha$ is the scaling factor for a submersed sphere and $\Sigma^2$ is the dimensionless resultant weight of the particles, determined in Eq.~9 in Ref.~\cite{vella2005cheerios}.  The capillary length, $L_c=\sqrt{\sigma/(\rho g)}$ , determines the length scale of the interfacial deflection and the Bond number $B_o=d^2/(4L_c^2)=\rho g d^2/(4\sigma)$ compares the relative importance of gravity and surface tension. At early times, the displacement between particles is small so that we can use the asymptotic form of the modified Bessel’s function: $K_1(x) \approx 1/x$ when $x \ll 1$. After using
this approximation in Eq.~\ref{eq3}, we Taylor expand to first order in $u(x)$ and take the continuum limit. (See full derivation in Supplementary Information.) We obtain:

\begin{equation}\label{eq4}
     \frac{\partial u}{\partial t} = - ( \frac{\sigma B_o^{5/2} \Sigma^2 L_c}{3\eta  \alpha} ) \frac{\partial^2 u}{\partial x^2} = - D \frac{\partial^2 u}{\partial x^2}
\end{equation}

Because all the terms inside the parentheses are positive, $D$ is also positive so that this expression has the form of a diffusion equation with a negative diffusion coefficient.  Therefore this system is unstable at all wavelengths; given a chance it will revert back to its unstretched initial condition. (The motion will be cut off when the separation between particles goes to zero.)  However, to do so involves a competition between the $k$-dependent relaxation and the pulling velocity, $V$, which affinely stretches the raft.

We can evaluate this competition by plugging in $u \propto e^{t/\tau}e^{ikx}$ to obtain a characteristic time scale  $\tau \propto k^{-2}$. We then extract a characteristic relaxation, or “healing”, velocity by rewriting the solution as  $e^{ik(x - ivt)}$ to obtain $v_{heal}= |iv| \approx \frac{1}{k \tau}$. This diffusive healing velocity depends on the wavelength of the cluster. Larger clusters naturally relax at a slower rate as, in one part of the cluster, the particles come together at the expense of the rest of the cluster where the particles move apart. By comparing the healing velocity, $v_{heal}$, with the velocity difference between adjacent particles due to the expanding metric of the pulling, $V/N_x$, we determine on what (small) scale the system looks relaxed and on what (larger) scale it looks as if it is still being pulled apart. By equating $v_{heal}\approx V/N_x$, we find the dependence on $V$ of the crossover or cluster-size wavevector $k_{cl}$ between these two regimes:

\begin{equation}\label{eq5}
k_{cl} \approx \frac{3
\eta\alpha} {\sigma B_o^{5/2} \Sigma^2 L_c}\frac{V}{N_x}.
\end{equation}

or, using $\langle\ell\rangle = 2\pi/k_{cl}$,

\begin{align}
\langle\ell\rangle &\approx 2\pi \frac{\sigma B_o^{5/2} \Sigma^2 L_c N_x}{3 \eta \alpha} V^{-1} \nonumber \\
&= \frac{\pi}{48} \frac{\Sigma^2}{\alpha} \frac{d^5 (\rho g)^2 N_x }{\sigma \eta}  V^{-1} 
\nonumber \\
&=(2 \pi N_x D)V^{-1}. \label{eq6}
\end{align}

In the second equation, we have expressed the result in terms of the experimental variables of the liquid and particle parameters. In the last equation, we express the result in terms of a single parameter, $D$, which is an experimentally measurable quantity as discussed below.

We first compare this result from the model, $\langle \ell \rangle=\frac{1}{a_{th}V^{b_{th}}}$, with the air-water data. The value of the exponent, $b_{th}=1$, is consistent with the experimental value: $b=1.1\pm0.2$. We also compare the magnitude of the prefactor in the model with that found in the experiment. We use an average value $d=600$ $\mu m$ and the value, $\Sigma^2/\alpha = 0.673$, determined in Ref. \cite{vella2005cheerios} for that value of $d$. We obtain  $a_{th}\approx 2.4 \times 10^3$ $sm^{-2}$. When compared with the fitted value $a = (1.2 \pm 0.4) \times 10^4$ $sm^{-2}$, the prefactor of the model only differs from the fitted value by roughly a factor of $5$. This is a surprisingly good agreement given that the model neglects all two-dimensional effects that are inherent in the experiment.

In Eq.~\ref{eq4}, $D$ is the ratio of the lateral capillary force to the Stokes’ drag, a quantity that can be directly measured in experiment. To obtain $D$, we track the trajectories of two particles at the air-liquid interface as they approach each other as described in Ref. \cite{vella2005cheerios, PhysRevE.83.051403}. We evaluate the effect of polydispersity by pairing large/large, large/small, and small/small particles. We averaged the measurements of $D$ under the assumption that large/small pairs are twice as likely to be found than large/large or small/small pairs in a well-mixed packing with equal numbers of large and small spheres. The results for water and the glycerol-water mixture are $D_{w} = (8.0 \pm 4.7) \times 10^{-7}$ $m^2 s^{-1}$ and $D_{gw} = (7.8 \pm 4.1) \times 10^{-8}$ $m^2 s^{-1}$ respectively.  These values are very close to the theoretical calculation for a pair of identical spheres with $d=600$ $\mu m$: $D = 7.60 \times 10^{-7}$ $m^2 s^{-1}$ and $D= 9.46 \times 10^{-8}$ $m^2 s^{-1}$  respectively. Thus the theoretical prediction is consistent with our measurements. However, with the one-dimensional model we are not able to account for how neighboring particles at different positions in the transverse direction affect the cluster size in the pulling direction.

The predicted crossover cluster length, $\langle \ell \rangle$, depends not only on velocity but also on fluid parameters,  $\eta$, $\sigma$ and $\rho$, which are explicit in the negative diffusion coefficient, $D$, and the size of the system, $N_x$. By changing the fluid, we can check the dependence of $\langle \ell \rangle$ on these parameters. 

Using our experimental measurements of the diffusion constant, $D$, for the two liquids, we scale $V$ with $(2 \pi N_x D_w)^{-1}$ and $(2\pi N_x D_{gw})^{-1}$ and find a good collapse between the water and glycerol-water mixtures, as shown in Fig.~\ref{fig3}B. In this data collapse, the dynamic viscosity contributes the most to the change in $D$. 

\section{Conclusions}

A particle raft floating on a liquid surface can be readily pulled apart to create an intriguingly intricate array of rifts separating condensed clusters of particles.  This failure mode is an accessible macroscopic analog of material failure that has counterparts at sizes ranging from the molecular scale of porous membranes up to the structure formation left behind by the expansion of the universe.  Failure is uniformly distributed within the raft and the size of the clusters formed in this process is controlled by the expansion velocity at which the raft is stretched. The observable size of the particles allows a thorough experimental investigation of the failure.

The different morphologies are caused by a single form of failure that is governed by expansion rate and not by a change in the characteristic mode of failure (e.g., brittle to ductile) as has previously been suggested~\cite{ARCINIAGA201136}. 
In this regard, it is similar to the situation of material breakup under uniaxial pulling near a rigidity transition~\cite{Driscoll10813} in which a zone of broken bonds can be tuned to extend over the entire width of the sample. In both cases, a single control parameter (pulling velocity in rafts and rigidity in networks) tunes the extent of the diffuse failure.
Neither situation fits naturally into the brittle/ductile dichotomy. 

In the rafts, the average cluster length of contiguous particles, $\langle \ell \rangle$, decreases monotonically with increasing $V$ until it reaches the single-particle cut-off. We have modeled this behavior by a one-dimensional chain with overdamped dynamics. This model produces a diffusion equation with a negative diffusion coefficient. We compare the healing velocity, that is the speed at which the particles relax, with the velocity difference between particles produced by the motion of the underlying fluid. The competition between these two velocities gives rise to a crossover length, which separates the scale at which particles coalesce into a cluster from the scale at which the particles become pulled apart. 

There remain many interesting features in the system to be understood. One topic to be investigated is how the overall dimensions of the system, $L_{x0}$ and $L_{y0}$, interact with the micro-crack formation. By concentrating on these rafts, where the microscopic rearrangements as well as the overall fracture dynamics can be assessed, the interaction with the edges can be addressed carefully. Thus, the dynamics and the dependence on system size and aspect ratio might be related to classical fracture mechanics. We also note that this phenomenon of micro-crack formation in rafts is reminiscent of other aspects of failure in the context of material processing. For example, the global shape of the rafts exhibits a change in Poisson’s ratio as a function of $V$, the pulling speed. At high $V$, the raft shows a near-zero Poisson’s ratio during the stretching. This occurs while most of the raft is still connected as a network and there are evenly distributed micro-cracks throughout the interior.  Such behavior is reminiscent of the formation of microstructure in some porous media, such as the expanded polytetrafluoroethylene (ePTFE)~\cite{gore1976process, kitamura1999formation}, which is not well understood. 

Although the raft is pulled along one direction, the clusters that are formed do not show a systematic orientation with respect to the flow. One might have expected that not only the rifts, but the cluster orientations, would show some aspect of the asymmetry of the dynamics. This is not the case.  There are important aspects of cluster formation that remain to be examined. In particular, it is of interest to explore the failure morphology of a raft pulled uniformly in a radial flow. In radial expansion, there is no single tensile-stress axis in the system. Such an expanding two-dimensional metric, as distinct from the one-dimensional pulling we have used here, would provide further insight into the situation of distributed failure in a fully three-dimensional system. Such an experiment would be more relevant to the situation of cosmological expansion. 

\section*{Author Contributions}
 K.-I. T. and S.R.N. conceived of the research, interpreted results, and wrote the paper. K.-I. T. performed experiments.

\section*{Conflicts of interest}
The authors declare that they have no competing interests.

\section*{Acknowledgements}
We thank M. Driscoll who made some of the first movies of these expanding rafts in our lab. We are also grateful to D. Hexner, J. Sethna, V. Vitelli, S. Lee and X. Cheng for enlightening discussions about how to think about this system. Funding: This work was supported primarily by the National Science Foundation (MRSEC program NSF-DMR 2011854). K.-I. T. was supported in part by Government Scholarship to Study Abroad by the Ministry of Education in Taiwan and by the Simons Foundation for the collaboration Cracking the Glass Problem Award \#348125.  



\balance


\providecommand*{\mcitethebibliography}{\thebibliography}
\csname @ifundefined\endcsname{endmcitethebibliography}
{\let\endmcitethebibliography\endthebibliography}{}

\bibliographystyle{rsc} 

\begin{mcitethebibliography}{30}
\providecommand*{\natexlab}[1]{#1}
\providecommand*{\mciteSetBstSublistMode}[1]{}
\providecommand*{\mciteSetBstMaxWidthForm}[2]{}
\providecommand*{\mciteBstWouldAddEndPuncttrue}
  {\def\EndOfBibitem{\unskip.}}
\providecommand*{\mciteBstWouldAddEndPunctfalse}
  {\let\EndOfBibitem\relax}
\providecommand*{\mciteSetBstMidEndSepPunct}[3]{}
\providecommand*{\mciteSetBstSublistLabelBeginEnd}[3]{}
\providecommand*{\EndOfBibitem}{}
\mciteSetBstSublistMode{f}
\mciteSetBstMaxWidthForm{subitem}
{(\emph{\alph{mcitesubitemcount}})}
\mciteSetBstSublistLabelBeginEnd{\mcitemaxwidthsubitemform\space}
{\relax}{\relax}

\bibitem[Fineberg and Marder(1999)]{FINEBERG19991}
J.~Fineberg and M.~Marder, \emph{Physics Reports}, 1999, \textbf{313},
  1--108\relax
\mciteBstWouldAddEndPuncttrue
\mciteSetBstMidEndSepPunct{\mcitedefaultmidpunct}
{\mcitedefaultendpunct}{\mcitedefaultseppunct}\relax
\EndOfBibitem
\bibitem[Bouchbinder \emph{et~al.}(2010)Bouchbinder, Fineberg, and
  Marder]{doi:10.1146/annurev-conmatphys-070909-104019}
E.~Bouchbinder, J.~Fineberg and M.~Marder, \emph{Annual Review of Condensed
  Matter Physics}, 2010, \textbf{1}, 371--395\relax
\mciteBstWouldAddEndPuncttrue
\mciteSetBstMidEndSepPunct{\mcitedefaultmidpunct}
{\mcitedefaultendpunct}{\mcitedefaultseppunct}\relax
\EndOfBibitem
\bibitem[Falk and Langer(2011)]{doi:10.1146/annurev-conmatphys-062910-140452}
M.~L. Falk and J.~Langer, \emph{Annual Review of Condensed Matter Physics},
  2011, \textbf{2}, 353--373\relax
\mciteBstWouldAddEndPuncttrue
\mciteSetBstMidEndSepPunct{\mcitedefaultmidpunct}
{\mcitedefaultendpunct}{\mcitedefaultseppunct}\relax
\EndOfBibitem
\bibitem[Nicolas \emph{et~al.}(2018)Nicolas, Ferrero, Martens, and
  Barrat]{RevModPhys.90.045006}
A.~Nicolas, E.~E. Ferrero, K.~Martens and J.-L. Barrat, \emph{Rev. Mod. Phys.},
  2018, \textbf{90}, 045006\relax
\mciteBstWouldAddEndPuncttrue
\mciteSetBstMidEndSepPunct{\mcitedefaultmidpunct}
{\mcitedefaultendpunct}{\mcitedefaultseppunct}\relax
\EndOfBibitem
\bibitem[Roux \emph{et~al.}(1988)Roux, Hansen, Herrmann, and Guyon]{Roux1988}
S.~Roux, A.~Hansen, H.~Herrmann and E.~Guyon, \emph{Journal of Statistical
  Physics}, 1988, \textbf{52}, 237--244\relax
\mciteBstWouldAddEndPuncttrue
\mciteSetBstMidEndSepPunct{\mcitedefaultmidpunct}
{\mcitedefaultendpunct}{\mcitedefaultseppunct}\relax
\EndOfBibitem
\bibitem[Malakhovsky and Michels(2007)]{PhysRevB.76.144201}
I.~Malakhovsky and M.~A.~J. Michels, \emph{Phys. Rev. B}, 2007, \textbf{76},
  144201\relax
\mciteBstWouldAddEndPuncttrue
\mciteSetBstMidEndSepPunct{\mcitedefaultmidpunct}
{\mcitedefaultendpunct}{\mcitedefaultseppunct}\relax
\EndOfBibitem
\bibitem[Alava \emph{et~al.}(2008)Alava, Nukala, and
  Zapperi]{PhysRevLett.100.055502}
M.~J. Alava, P.~K. V.~V. Nukala and S.~Zapperi, \emph{Phys. Rev. Lett.}, 2008,
  \textbf{100}, 055502\relax
\mciteBstWouldAddEndPuncttrue
\mciteSetBstMidEndSepPunct{\mcitedefaultmidpunct}
{\mcitedefaultendpunct}{\mcitedefaultseppunct}\relax
\EndOfBibitem
\bibitem[Shekhawat \emph{et~al.}(2013)Shekhawat, Zapperi, and
  Sethna]{PhysRevLett.110.185505}
A.~Shekhawat, S.~Zapperi and J.~P. Sethna, \emph{Phys. Rev. Lett.}, 2013,
  \textbf{110}, 185505\relax
\mciteBstWouldAddEndPuncttrue
\mciteSetBstMidEndSepPunct{\mcitedefaultmidpunct}
{\mcitedefaultendpunct}{\mcitedefaultseppunct}\relax
\EndOfBibitem
\bibitem[Driscoll \emph{et~al.}(2016)Driscoll, Chen, Beuman, Ulrich, Nagel, and
  Vitelli]{Driscoll10813}
M.~M. Driscoll, B.~G.-g. Chen, T.~H. Beuman, S.~Ulrich, S.~R. Nagel and
  V.~Vitelli, \emph{Proceedings of the National Academy of Sciences}, 2016,
  \textbf{113}, 10813--10817\relax
\mciteBstWouldAddEndPuncttrue
\mciteSetBstMidEndSepPunct{\mcitedefaultmidpunct}
{\mcitedefaultendpunct}{\mcitedefaultseppunct}\relax
\EndOfBibitem
\bibitem[Varshney \emph{et~al.}(2012)Varshney, Sane, Ghosh, and
  Bhattacharya]{PhysRevE.86.031402}
A.~Varshney, A.~Sane, S.~Ghosh and S.~Bhattacharya, \emph{Phys. Rev. E}, 2012,
  \textbf{86}, 031402\relax
\mciteBstWouldAddEndPuncttrue
\mciteSetBstMidEndSepPunct{\mcitedefaultmidpunct}
{\mcitedefaultendpunct}{\mcitedefaultseppunct}\relax
\EndOfBibitem
\bibitem[Jambon-Puillet \emph{et~al.}(2017)Jambon-Puillet, Josserand, and
  Proti\`ere]{PhysRevMaterials.1.042601}
E.~Jambon-Puillet, C.~Josserand and S.~Proti\`ere, \emph{Phys. Rev. Materials},
  2017, \textbf{1}, 042601\relax
\mciteBstWouldAddEndPuncttrue
\mciteSetBstMidEndSepPunct{\mcitedefaultmidpunct}
{\mcitedefaultendpunct}{\mcitedefaultseppunct}\relax
\EndOfBibitem
\bibitem[Cicuta and Vella(2009)]{PhysRevLett.102.138302}
P.~Cicuta and D.~Vella, \emph{Phys. Rev. Lett.}, 2009, \textbf{102},
  138302\relax
\mciteBstWouldAddEndPuncttrue
\mciteSetBstMidEndSepPunct{\mcitedefaultmidpunct}
{\mcitedefaultendpunct}{\mcitedefaultseppunct}\relax
\EndOfBibitem
\bibitem[Vella \emph{et~al.}(2004)Vella, Aussillous, and Mahadevan]{Vella_2004}
D.~Vella, P.~Aussillous and L.~Mahadevan, \emph{Europhysics Letters ({EPL})},
  2004, \textbf{68}, 212--218\relax
\mciteBstWouldAddEndPuncttrue
\mciteSetBstMidEndSepPunct{\mcitedefaultmidpunct}
{\mcitedefaultendpunct}{\mcitedefaultseppunct}\relax
\EndOfBibitem
\bibitem[Razavi \emph{et~al.}(2015)Razavi, Cao, Lin, Lee, Tu, and
  Kretzschmar]{doi:10.1021/acs.langmuir.5b01652}
S.~Razavi, K.~D. Cao, B.~Lin, K.~Y.~C. Lee, R.~S. Tu and I.~Kretzschmar,
  \emph{Langmuir}, 2015, \textbf{31}, 7764--7775\relax
\mciteBstWouldAddEndPuncttrue
\mciteSetBstMidEndSepPunct{\mcitedefaultmidpunct}
{\mcitedefaultendpunct}{\mcitedefaultseppunct}\relax
\EndOfBibitem
\bibitem[Bandi \emph{et~al.}(2011)Bandi, Tallinen, and Mahadevan]{Bandi_2011}
M.~M. Bandi, T.~Tallinen and L.~Mahadevan, \emph{{EPL} (Europhysics Letters)},
  2011, \textbf{96}, 36008\relax
\mciteBstWouldAddEndPuncttrue
\mciteSetBstMidEndSepPunct{\mcitedefaultmidpunct}
{\mcitedefaultendpunct}{\mcitedefaultseppunct}\relax
\EndOfBibitem
\bibitem[Peco \emph{et~al.}(2017)Peco, Chen, Liu, Bandi, Dolbow, and
  Fried]{peco2017influence}
C.~Peco, W.~Chen, Y.~Liu, M.~Bandi, J.~E. Dolbow and E.~Fried, \emph{Soft
  matter}, 2017, \textbf{13}, 5832--5841\relax
\mciteBstWouldAddEndPuncttrue
\mciteSetBstMidEndSepPunct{\mcitedefaultmidpunct}
{\mcitedefaultendpunct}{\mcitedefaultseppunct}\relax
\EndOfBibitem
\bibitem[Kim \emph{et~al.}(2019)Kim, Rendos, Ganesh, and Brown]{KIM201954}
B.~L. Kim, A.~Rendos, P.~Ganesh and K.~A. Brown, \emph{Colloid and Interface
  Science Communications}, 2019, \textbf{28}, 54--59\relax
\mciteBstWouldAddEndPuncttrue
\mciteSetBstMidEndSepPunct{\mcitedefaultmidpunct}
{\mcitedefaultendpunct}{\mcitedefaultseppunct}\relax
\EndOfBibitem
\bibitem[Xiao \emph{et~al.}(2020)Xiao, Ivancic, and Durian]{D0SM00839G}
H.~Xiao, R.~J.~S. Ivancic and D.~J. Durian, \emph{Soft Matter}, 2020,
  \textbf{16}, 8226--8236\relax
\mciteBstWouldAddEndPuncttrue
\mciteSetBstMidEndSepPunct{\mcitedefaultmidpunct}
{\mcitedefaultendpunct}{\mcitedefaultseppunct}\relax
\EndOfBibitem
\bibitem[Arciniaga \emph{et~al.}(2011)Arciniaga, Kuo, and
  Dennin]{ARCINIAGA201136}
M.~Arciniaga, C.-C. Kuo and M.~Dennin, \emph{Colloids and Surfaces A:
  Physicochemical and Engineering Aspects}, 2011, \textbf{382}, 36 -- 41\relax
\mciteBstWouldAddEndPuncttrue
\mciteSetBstMidEndSepPunct{\mcitedefaultmidpunct}
{\mcitedefaultendpunct}{\mcitedefaultseppunct}\relax
\EndOfBibitem
\bibitem[Wang \emph{et~al.}(2015)Wang, Kanjanaboos, McBride, Barry, Lin, and
  Jaeger]{C4FD00243A}
Y.~Wang, P.~Kanjanaboos, S.~P. McBride, E.~Barry, X.-M. Lin and H.~M. Jaeger,
  \emph{Faraday Discuss.}, 2015, \textbf{181}, 325--338\relax
\mciteBstWouldAddEndPuncttrue
\mciteSetBstMidEndSepPunct{\mcitedefaultmidpunct}
{\mcitedefaultendpunct}{\mcitedefaultseppunct}\relax
\EndOfBibitem
\bibitem[Lu \emph{et~al.}(2010)Lu, Suo, and Vlassak]{LU20101679}
N.~Lu, Z.~Suo and J.~J. Vlassak, \emph{Acta Materialia}, 2010, \textbf{58},
  1679--1687\relax
\mciteBstWouldAddEndPuncttrue
\mciteSetBstMidEndSepPunct{\mcitedefaultmidpunct}
{\mcitedefaultendpunct}{\mcitedefaultseppunct}\relax
\EndOfBibitem
\bibitem[Skjeltorp and Meakin(1988)]{skjeltorp1988fracture}
A.~Skjeltorp and P.~Meakin, \emph{Nature}, 1988, \textbf{335}, 424--426\relax
\mciteBstWouldAddEndPuncttrue
\mciteSetBstMidEndSepPunct{\mcitedefaultmidpunct}
{\mcitedefaultendpunct}{\mcitedefaultseppunct}\relax
\EndOfBibitem
\bibitem[Routh(2013)]{Routh_2013}
A.~F. Routh, \emph{Reports on Progress in Physics}, 2013, \textbf{76},
  046603\relax
\mciteBstWouldAddEndPuncttrue
\mciteSetBstMidEndSepPunct{\mcitedefaultmidpunct}
{\mcitedefaultendpunct}{\mcitedefaultseppunct}\relax
\EndOfBibitem
\bibitem[Paunov \emph{et~al.}(1993)Paunov, Kralchevsky, Denkov, and
  Nagayama]{paunov1993lateral}
V.~Paunov, P.~Kralchevsky, N.~Denkov and K.~Nagayama, \emph{Journal of colloid
  and interface science}, 1993, \textbf{157}, 100--112\relax
\mciteBstWouldAddEndPuncttrue
\mciteSetBstMidEndSepPunct{\mcitedefaultmidpunct}
{\mcitedefaultendpunct}{\mcitedefaultseppunct}\relax
\EndOfBibitem
\bibitem[Vella and Mahadevan(2005)]{vella2005cheerios}
D.~Vella and L.~Mahadevan, \emph{American journal of physics}, 2005,
  \textbf{73}, 817--825\relax
\mciteBstWouldAddEndPuncttrue
\mciteSetBstMidEndSepPunct{\mcitedefaultmidpunct}
{\mcitedefaultendpunct}{\mcitedefaultseppunct}\relax
\EndOfBibitem
\bibitem[Dalbe \emph{et~al.}(2011)Dalbe, Cosic, Berhanu, and
  Kudrolli]{PhysRevE.83.051403}
M.-J. Dalbe, D.~Cosic, M.~Berhanu and A.~Kudrolli, \emph{Phys. Rev. E}, 2011,
  \textbf{83}, 051403\relax
\mciteBstWouldAddEndPuncttrue
\mciteSetBstMidEndSepPunct{\mcitedefaultmidpunct}
{\mcitedefaultendpunct}{\mcitedefaultseppunct}\relax
\EndOfBibitem
\bibitem[George~B. and Hans~J.(2005)]{18717320050101}
A.~George~B. and W.~Hans~J., \emph{Mathematical Methods For Physicists
  International Student Edition.}, Academic Press, 2005, vol. 6th ed. George B.
  Arfken, Hans J. Weber, pp. 79--83\relax
\mciteBstWouldAddEndPuncttrue
\mciteSetBstMidEndSepPunct{\mcitedefaultmidpunct}
{\mcitedefaultendpunct}{\mcitedefaultseppunct}\relax
\EndOfBibitem
\bibitem[Kravtsov and Borgani(2012)]{doi:10.1146/annurev-astro-081811-125502}
A.~V. Kravtsov and S.~Borgani, \emph{Annual Review of Astronomy and
  Astrophysics}, 2012, \textbf{50}, 353--409\relax
\mciteBstWouldAddEndPuncttrue
\mciteSetBstMidEndSepPunct{\mcitedefaultmidpunct}
{\mcitedefaultendpunct}{\mcitedefaultseppunct}\relax
\EndOfBibitem
\bibitem[Gore(1976)]{gore1976process}
R.~W. Gore, \emph{Process for producing porous products}, 1976, US Patent
  3,953,566\relax
\mciteBstWouldAddEndPuncttrue
\mciteSetBstMidEndSepPunct{\mcitedefaultmidpunct}
{\mcitedefaultendpunct}{\mcitedefaultseppunct}\relax
\EndOfBibitem
\bibitem[Kitamura \emph{et~al.}(1999)Kitamura, Kurumada, Tanigaki, Ohshima, and
  Kanazawa]{kitamura1999formation}
T.~Kitamura, K.-I. Kurumada, M.~Tanigaki, M.~Ohshima and S.-I. Kanazawa,
  \emph{Polymer Engineering \& Science}, 1999, \textbf{39}, 2256--2263\relax
\mciteBstWouldAddEndPuncttrue
\mciteSetBstMidEndSepPunct{\mcitedefaultmidpunct}
{\mcitedefaultendpunct}{\mcitedefaultseppunct}\relax
\EndOfBibitem
\end{mcitethebibliography}

\end{document}